\documentclass[fleqn,usenatbib]{mnras}

\usepackage{newtxtext,newtxmath}
\usepackage[T1]{fontenc}

\DeclareRobustCommand{\VAN}[3]{#2}
\let\VANthebibliography\thebibliography
\def\thebibliography{\DeclareRobustCommand{\VAN}[3]{##3}\VANthebibliography}

\usepackage{graphicx}	% Including figure files
\usepackage{amsmath}	% Advanced maths commands

\title[The study of AF~And. I.]{The study of AF~And. I. Hydrogen abundance constraints and dynamically-consistent wind model of the hot state}

\author[E. O. Dedov et al.]{
E. O. Dedov,$^{1}$\thanks{E-mail: evdedov@sao.ru (EOD)}
A. E. Kostenkov,$^{1}$
A. S. Vinokurov$^{1}$
and A. N. Sarkisyan$^{1}$
\\
$^{1}$Special Astrophysical Observatory, Nizhnij Arkhyz 369167, Russia\\
}

\date{Accepted XXX. Received YYY; in original form ZZZ}

\pubyear{\the\year{}}

\begin{document}
\label{firstpage}
\pagerange{\pageref{firstpage}--\pageref{lastpage}}
\maketitle

% Abstract of the paper
\begin{abstract}
This paper starts a series of works devoted to an in-depth study of luminous blue variable AF~And in the M\,31 galaxy. Here we present the results of non-LTE modelling of the star in a hot state with \textsc{cmfgen} code. In addition to the $\beta$-velocity law, widely used in the analysis of stellar atmospheres, we used a velocity distribution obtained from the balance of radial forces in the wind, including centrifugal forces for several models. Such a dynamically consistent approach reduces the parameter space during modelling and, in addition, allows us to estimate the current stellar mass whose inner hydrostatic layers are obscured by a dense wind that contaminates the absorption profiles. Based on our modelling we estimate the parameters of AF~And as follows: luminosity $L \approx 6.5\times10^5~\text{L}_\odot$, mass loss rate $\Dot{M} \approx 3\times10^{-5}~\text{M}_\odot\,\text{yr}^{-1}$, terminal velocity $v_\infty \approx 300~\text{km\,s}^{-1}$ and effective temperature $T_{\rm eff} \approx 17~\text{kK}$. modelling of the hot state of AF~And is of particular interest due to the possibility to place constraints on the ratio of hydrogen and helium abundances for further use in constructing models of the cold state of this star. We conclude that the current hydrogen mass fraction in the wind lies in the range from $20$ to $30$~per~cent. Self-consistent modelling gives the current mass $M_*$ estimation in the range $15.6 - 17.7~\text{M}_\odot$. By employing evolutionary tracks, we were able to estimate the initial stellar mass as $M_{*,~\rm{init}} \approx 50~\text{M}_\odot$.
\end{abstract}

% Select between one and six entries from the list of approved keywords.
% Don't make up new ones.
\begin{keywords}
stars: fundamental parameters -- stars: mass-loss -- stars: variables: S Doradus -- stars: individual: AF And
\end{keywords}

%%%%%%%%%%%%%%%%%%%%%%%%%%%%%%%%%%%%%%%%%%%%%%%%%%

%%%%%%%%%%%%%%%%% BODY OF PAPER %%%%%%%%%%%%%%%%%%

\section{Introduction}
\label{sect:intro}

Luminous blue variables (LBVs) represent a relatively rare class of massive stars ($M_{\rm init} \gtrsim 25~\text{M}_\odot$) with extremely high luminosities ($L \gtrsim 10^5~\text{L}_\odot$) \citep{Humphreys2014, Humphreys2017, Smith2026}. The intense interest in these objects stems, in part, from their influence on the evolution of the surrounding interstellar medium via strong stellar winds. According to classical concepts, a massive star is usually called an LBV while undergoing a relatively short evolutionary stage ($10^3 - 10^5$~years) \citep{Humphreys1991}. This period of stellar evolution is important because it is believed to be a transitional stage between an O star and the hydrogen-poor stage of Wolf-Rayet (WR) stars \citep{Conti1975, Conti1984, Langer1994}. However, modern studies of supernovae progenitors and massive stars evolutionary scenarios show that LBVs can be progenitors of Type IIn supernovae \citep{Gal-Yam2009, Ustamujic2021} or IIb supernovae \citep{Kotak2006, Groh2013}.
As an LBV, a star exhibits significant photometric (up to $\Delta m = 2.5$~mag in the $V$ band) and spectral variability, often referred to as S~Dor variability \citep{Weis2020, Smith2026}. Photometric variability is accompanied by changes in the spectrum: at near-minimum brightness, the spectrum of a B-supergiant or Of/late–WN stars is observed (the `hot' state), and as the maximum brightness is approached, the spectrum changes to a spectrum similar to an A-F supergiant (the `cold' state) \citep{Humphreys1994, Humphreys2014}. It is worth noting that LBVs can also exhibit irregular high-amplitude outbursts ($\Delta m \gtrsim 2.5$~mag) \citet{Smith2026}, similar to those observed for $\eta$~Car \citep{SmithLi2011} and P~Cygni \citep{Israelian1999}. The characteristic durations of variability cycles are not precisely known. There are objects for which both `hot' and `cold' states have been observed (e.g., AG~Car \citep{Groh2009AGI, Groh2011}, R\,127 \citep{Walborn2008}, R\,71 \citep{Lennon1993, Mehner2013}), as well as objects that remain at minimum brightness for hundreds of years (`dormant' LBVs, e.g., P~Cygni \citep{Elliott2022}, Var\,2, see references in \citet{Kostenkov2026}).

To determine the physical mechanisms of these objects' variability, as well as their evolutionary status, it is important to obtain values of their fundamental parameters (luminosity $L$, effective temperature $T_{\rm eff}$, mass loss rate $\Dot{M}$, current mass $M_*$). For massive stars in advanced stages of evolution, this task is complicated by the presence of an extended atmosphere, the conditions of which are far from LTE \citep{Hillier1997, Vink1999}. Currently, the most accurate method for determining stellar parameters is the construction of non-LTE models of stellar atmospheres. The standard approach does not allow mass estimation directly from the atmospheric model, since the presence of a dense extended atmosphere masks the deep photospheric absorption lines used for mass estimation. This problem can be solved by constructing self-consistent hydrodynamic atmospheric models in which velocity distributions are calculated based on the physical properties of the wind rather than postulated in the form of the widely used $\beta$-law. Significant progress in constructing such models has been achieved over the past decades \citep{MullerVink2008, Sander2017, Gormaz-Matamala2021, Gonzalez-Tora2025a, Gonzalez-Tora2025b}. The results obtained by \cite{Sabhahit2025} for R144 binary system consisting of WN5/6h+WN6/7h are of particular interest due to the availability of independent stellar mass measurements. The reduction of the number of free parameters allowed the authors to derive clumping law parameters (see below for details) more reliably.

Due to the demands of this method on both the quality of spectral data and the computational time, a significant number of models have been constructed to date only for OB super- and hypergiants \citep{Aschenbrenner2026, Bernini-Peron2026}, and WR stars \citep{Grafener2005, Lefever2026}, while similar studies for LBV-type winds are scarce, especially those covering the `cold' state.

Non-LTE models of extended atmospheres\footnote{These studies used standard $\beta$-law for the wind velocity distribution.} have been applied to estimate the parameters of several recently discovered high-luminosity variable stars with optical spectra, that resembling those of LBVs in their cool state \citep{Solovyeva2021, Solovyeva2023}. It was found that \ion{Fe}{ii} to hydrogen lines ratio are well reproduced at low hydrogen abundances in the wind (approximately 20~per~cent by mass fraction, \citealt{Solovyeva2021,Solovyeva2022}), which is more typical of WR stars than LBVs \citep{Davies2010}. It may be possible that these results were obtained due to the degeneracy of the hydrogen abundance~--~mass-loss rate parameters in stellar atmospheres with effective temperatures of approximately 10~kK. Such objects typically demonstrate the absence of important diagnostic He lines which makes it impossible to directly constrain the H/He abundance ratio. \cite{Hillier1998} discussed this issue in details in context of the study of the Galactic supergiant HDE~316285 and concluded that the observed spectrum can be reproduced by models in the wide range of H/He ratios, although the authors state that a $\text{H/He}\approx1.5$ by number\footnote{corresponds to hydrogen abundance of approximately 27~per~cent by mass fraction} is preferable according to overall quality of describing metals line profiles. The degeneracy problem was also discussed in \citet{Najarro2009}, as applied to the stars of Quintuplet cluster. In that paper the degeneracy of the parameters was resolved using the infrared data, particularly, \ion{He}{i}\,$\lambda2.112~\mu\text{m}$. We propose to resolve the `hydrogen abundance~--~mass-loss rate' coupling in `cold' LBVs atmospheres with a different approach. The hot state exhibits strong lines of both \ion{H}{i} and \ion{He}{i} (\ion{He}{ii} lines may also be observed in some cases) that would allow for direct estimation of H/He ratio.

One example of an object for which such a study is feasible is the confirmed LBV star AF~And in the M\,31 galaxy. The star was discovered about 100 years ago \citep{Luyten1927} and has since exhibited noticeable photometric variability on the scale of decades, including irregular outbursts of $\Delta m \approx 1.5$~mag \citep{Joshi2019}.

The spectroscopy had shown the transition of the spectrum from the `cold' state with numerous \ion{Fe}{ii} and [\ion{Fe}{ii}] emission lines \citep{Gallagher1981} towards the `hot' Of/late WN-like state \citep{Humphreys2014, Humphreys2017} resembling P~Cygni \citep{Najarro1997} and MN\,112 \citep{Kostenkov2020}. \citet{Valeev2015} had performed a radiative transfer modelling of AF~And in its `hot' state, but due to a low spectral resolution and signal-to-noise ratio, authors could only constrain several fundamental parameters.

This study marks the beginning of a series of papers devoted to AF~And. Our goal is to obtain estimates of the hydrogen abundance in the wind, the current mass, and the dynamical properties of the wind based on data available for the `hot' state, for subsequent use in modelling the `cold' state. The paper is organized as follows: Section \ref{sect:obs} describes the observational data used; Section \ref{sect:result} describes the process of constructing a non-LTE model to determine the star's fundamental parameters and gives details on the construction of self-consistent hydrodynamic models to obtain the star's current mass and refine other fundamental parameters. Section \ref{sect:discussion} contains a discussion of the results obtained, namely, the evolutionary status of AF~And and the influence of wind inhomogeneities on its properties. Section \ref{sect:conclusion} concludes the present study.

\section{Observational data}
\label{sect:obs}

We used the spectral data published in \citet{Humphreys2014}, that were obtained with MMT/HECTOSPEC with 600gpm grating (combined spectral range of B and R channels $3550 - 8050~\text{\AA}$, spectral resolution in the vicinity of H\,$\alpha$ line $\approx 2.1~\text{\AA}$) on 2010 October 10\footnote{The spectra may be downloaded from R. Humphreys' Eta Car Project website: \url{https://etacar.umn.edu/LuminousStars/M31M33/M31stars.html}}. Subsequent observations in 2013 and 2014 showed relative stability of the star's state \citep{Humphreys2014, Humphreys2017} and only the data from 2017 indicate the transition of AF~And to a cold state.

The optical spectrum in hot state contains noticeable emission lines of the Balmer series, numerous \ion{He}{i}, \ion{N}{ii} and \ion{Fe}{iii} lines, some of \ion{Si}{ii-iv} lines and weak \ion{C}{ii} lines were also detected. Many of the listed lines have a P~Cyg profile. The \ion{H}{i} lines are contaminated by the emission of the nebula, which also contributes with other nebular emission lines such as [\ion{N}{ii}], [\ion{S}{ii}]. Observed spectrum with line identification is presented on the figure \ref{fig:fullspec}.

We carried out additional observations of AF~And with the 6-m telescope of SAO RAS (BTA) equipped with SCORPIO-2 focal reducer \citep{Scorpio2011} on 2023 November 7. We had used VPGH1200@540 grating with $0.5^{\prime\prime}$ slit (spectral range $3550 - 8050~\text{\AA}$, spectral resolution in the vicinity of H\,$\alpha$ line $\approx 2.5~\text{\AA}$). With this additional data we were able to account for the presence of the projected nebula (see section \ref{sect:result}). Data reduction was performed with standard routines of \textsc{midas} environment. The spectrum of the nebula was extracted with \textsc{spextra} software \citep{Spextra2017} with the extraction region width of $8^{\prime\prime}$. The extraction region is located $8^{\prime\prime}$ west of the star.

Quasi-simultaneous imaging data of AF~And to correct for slit losses was found in CFHT archival data. $g$ and $r$ band frames were obtained on 2018 October 08 and 2010 October 09, respectively. Data processing included standard procedures of bias, flat-field corrections and image stacking. The magnitudes measured by aperture photometry using local standards yield $g = (16.981 \pm 0.014)$~mag, $r = (16.957 \pm 0.011)$~mag in ABMag system.

\section{Methods and results}
\label{sect:result}

\textsc{cmfgen} radiative-transfer code \citep{HillierMiller1998} used for the wind modelling does not solve the equation of motion for the wind, so the velocity distribution must be specified a priori. In the first stage of the study, we construct models using the $\beta$-law \ref{betalaw}, that then used to construct dynamically consistent wind models, in which the velocity distribution is determined by the balance of radial forces and, therefore, has no free parameters.

\subsection{$\beta$-law based models}
\label{sect:modelling}
\textsc{cmfgen} code solves the radiative transfer equation in a conjunction with statistical equilibrium equations in spherical geometry. \textsc{cmfgen} is widely used in constructing models of extended atmospheres of OB \citep{Martins2005, Crowther2006}, WR \citep{Bresolin2002, Crowther2007}, and LBV \citep{Najarro2001, Groh2009AGI} stars. The model is defined by a set of basic parameters, such as the luminosity $L$, the hydrostatic radius $R_*$\footnote{The hydrostatic radius is defined as the inner point of the atmosphere at which $\tau_{\rm Ross} \gtrsim 20$}, the mass-loss rate $\Dot{M}$, and individual abundances of chemical elements $X_i$. The wind velocity distribution is parameterized as the $\beta$-law \citep{Hillier2003}
\begin{equation}\label{betalaw}
    v(r)=\frac{v_0 + (v_{\infty} - v_0)(1 - R_* / r)^\beta}{1 + (v_0/v_\text{core})\exp[{(R_* - r)/h_\text{eff}}]},
\end{equation}

where $\beta$ is an exponent characterizing the efficiency of wind acceleration, $v_{\rm core}$ is the velocity at radius $R_*$, $v_0$ is the velocity in the intermediate zone between the inner atmosphere and the wind itself, $h$ is the effective height of the isothermal atmosphere, and $v_{\infty}$ is the terminal wind velocity.

Of all the velocity law parameters, the observed spectrum allows us to reliably estimate only the terminal velocity. We used for this purpose the forbidden line [\ion{N}{ii}]\,$\lambda5755$ and the positions of the absorption components of the P~Cyg profiles of the \ion{H}{i}. The value of $\beta$ is not strictly defined, and various studies report $\beta \sim 1 - 4$ \citep{Najarro1997, Hillier2003, Maryeva2012, Kostenkov2020}. This parameter can also be estimated from the positions of the absorption components of the P~Cyg line profiles of different ions, but this approach is less reliable due to the degeneracy between the acceleration and terminal velocity parameters, since both affect absorption velocity. As the most suitable value, $\beta = 2$ was chosen and fixed to reduce the number of free parameters. The remaining parameters of the velocity law were fixed as follows: $v_0 = 15~\text{km\,s}^{-1}$ to match the assumed speed of sound in the transitional part of the wind ($v_{\rm sonic} = 15~\text{km\,s}^{-1}$), $v_{\rm core}$ was chosen such that optical depth at $R_*$ was in the range of $20 \lesssim \tau_{\rm Ross} \lesssim 100$ when constructing the models. $h$ was chosen equal to $0.1 R_*$ as it optimally describe the hydrostatic structure \citep{Kostenkov2026}. The form of the adopted velocity law is demonstrated on the figure \ref{fig:vellaw}.

The set of emission lines in the AF~And spectrum allows us to estimate the temperature by combining several criteria. First, the temperature can be estimated based on the \ion{Si}{iii}/\ion{Si}{ii} line intensity ratio. Second, the depth of the absorption components of the Balmer series lines are also sensitive to temperature, since temperature changes the transparency of the Lyman continuum. Finally, the spectrum contains temperature-sensitive \ion{Fe}{iii}, \ion{N}{ii} and \ion{He}{i} lines, the intensities of which also limits the temperature.

It is evident from both observational and theoretical sources that winds of massive stars are inhomogeneous. Observational phenomena include e.g. the symmetric profiles of X-ray emission lines \citep{Oskinova2006}, stochastic linear polarization of emission lines \citep{Lupie1987}, strength of electron scattering wings of emission lines in comparison with the associated line intensity \citep{Hillier1991}. Theoretical investigations are usually conducted under several physical approximations, such as LTE or Sobolev method (e.g. \citealt{Sundqvist2018, Debnath2024}) due to the requirement of the application of complex and computationally demanding 2D or 3D geometry in addition to the necessity of taking into account time-dependent effects. This makes it difficult to directly incorporate the results of the simulations of exact density structures in 1D steady-state models constructed with the co-moving frame method. Consequently, simplified approaches known as micro- and macrocluping are generally adopted in such cases. \textsc{cmfgen} accounts for microscopic wind inhomogeneities (assuming empty space between clumps of matter) in the form of a distribution of the volume filling factor $f$. It is defined as the ratio between the density of clumped matter and uniformly distributed matter. The distribution of $f$ in the velocity field is represented by the law \citep{Najarro2011}
\begin{equation}\label{clump_param}
f(r)=f_1 + (1 - f_1)\exp\left(\frac{-v(r)}{f_2}\right) + (1 - f_1)\exp\left(\frac{v(r) - v_\infty}{f_3}\right)
\end{equation} 
where $f_1$ is the minimal value of the filling factor, $f_2$ and $f_3$ are the velocity values at which the filling factor increases and decreases, respectively. When $f_3 = 0$, the last term in the expression is ignored and $f(r)$ reduces to a simpler form, widely used in the analysis of extended atmospheres \citep{Hillier2003, Groh2009AGI, Maryeva2022}.

The parameters of the filling factor distribution were selected as follows. $f_1$ can be estimated from the intensity of the scattering wings in optical recombination lines, since the line intensities are determined by the square of the density, and for the wings the dependence on the density is linear, which means that a joint selection of $f_1$ and $\Dot{M}$ can provide an optimal description of these spectral features. The typical choice of $f_1$ lies in the range of $f_1 = 0.05-0.5$ \citep{Hillier1999clump, Hillier2003}. The shape of the $f(r)$ distribution, specified by the parameters $f_2$ and $f_3$, was chosen based on the results of the studies of the O4I(n)f star $\zeta$~Pup and time-dependent radiative-hydrodynamical modelling since we have not found any data on the behaviour of inhomogeneities for P~Cygni-like LBVs. \citet{Puls2006} and \citet{Najarro2011} showed that the minimum of the $f(r)$ function for $\zeta$~Pup occurs at $r = 2~R_*$. Several studies indicate that $f(r)$ reaches its minimum further away, at distances of $10-20~R_*$ \citep{Runacres2007, RubioDiez2022}. The choice of the values of $f(r)$ distribution parameters was based on the ratio between the intensities of \ion{He}{i} emission lines. We had computed approximately 100 models with different $f_1$ values and clumping onset locations, controlled by $f_2$ and $f_3$. The relation between \ion{He}{i}\,$\lambda5015,5876,6678$ lines is particularly sensitive to the shape of the $f(r)$ distribution (see e.g. \citealt{Kostenkov2026}). The investigation of the different clumping laws showed that $f_1=0.25, f_2 = 10~\text{km\,s}^{-1}, f_3 = 40~\text{km\,s}^{-1}$ values provide the relation between the intensities of the named \ion{He}{i} lines close to the observed one.

The metal abundances $X_i$ were fixed at solar values in accordance with the metallicity of the M\,31 galaxy \citep{Zurita2012}, and the N and C abundances were estimated from the \ion{N}{ii} and \ion{C}{ii} lines present in the spectrum of AF~And. The spectrum does not contain noticeable O lines, so we took $X_O = 0.01~X_{O,~\odot}$ as a typical value of oxygen abundance \citep{Groh2009AGI}. For nitrogen and carbon we adopt $X_N = 6.8~X_{N,~\odot}$ and $X_C = 0.1~X_{C,~\odot}$, respectively\footnote{Solar abundances $X_{i,~\odot}$ are taken from \citet{Asplund2009}.}. Our models include \ion{H}{i}, \ion{He}{i}, \ion{C}{i-iii}, \ion{N}{i-iii}, \ion{O}{i-iii}, \ion{Ne}{ii-iv}, \ion{Na}{i-ii}, \ion{Si}{ii-iv}, \ion{P}{iii-iv}, \ion{S}{iii-iv}, \ion{Ar}{iii-iv}, \ion{Cr}{ii-iv}, \ion{Fe}{ii-iv}, \ion{Ni}{ii-iv}, \ion{Co}{ii-iv} ions with total number of bound-bound transitions of $\approx 1.2 \times 10^5$ to ensure a realistic radiation field.

The luminosity was estimated by fitting the photometric data of the CFHT telescope to a model spectrum. For adopted distance $(785 \pm 25)~\text{kpc}$ based on the results of TRGB method by \citet{McConnachie2005} we obtained the value $L = 6.5\times10^5~\text{L}_\odot$ and simultaneously $A_V = 0.7$~mag with the model spectra fitting of the observed $g$ and $r$ fluxes. To check this value, we estimate $A_V$ with Balmer decrement of the surrounding nebula with our SCORPIO-2 data. The measurements were performed for several regions in the vicinity of AF~And and yielded $A_V = (0.71 \pm 0.05)$~mag assuming case B photoionization \citep{Osterbrock2006} for $R_V=3.1$.

The value $\Dot{M}$ was determined by fitting the intensities of the strong lines of \ion{H}{i} and \ion{He}{i}. However, the obtained estimate is significantly affected by the H/He abundance ratio. For this reason, we constructed several series of models with different hydrogen abundances to obtain a more reliable determination of the fundamental parameters.

\subsection{Hydrogen abundance}
\label{sect:hydrogen}

\begin{figure*}
    \centering
    \includegraphics[width=1.0\linewidth]{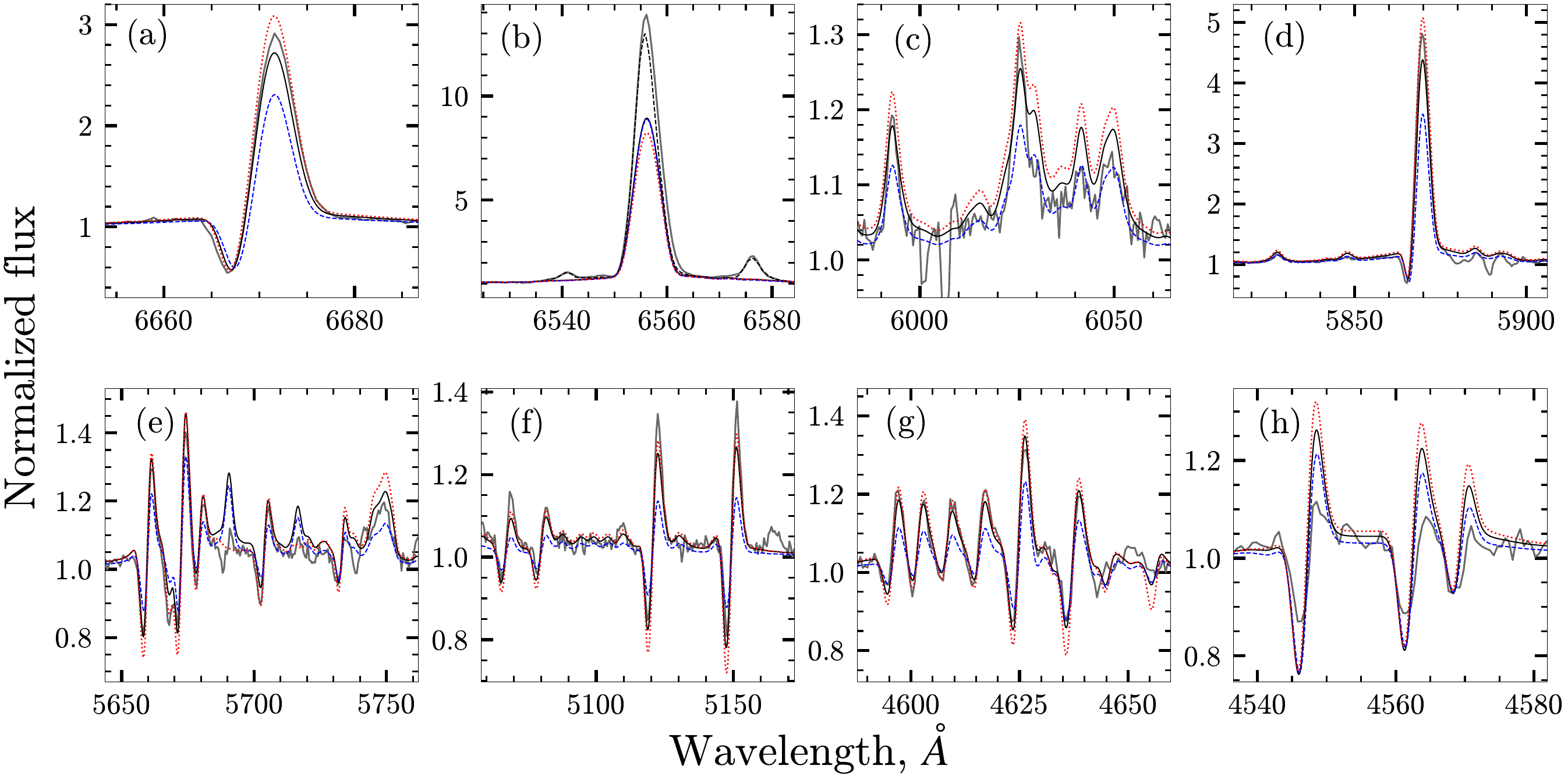}
    \caption{Comparison of the line profiles for several chosen lines in $\beta$-law based models with different hydrogen abundance. The normalized spectrum of AF~And is shown by grey solid line, low-hydrogen model with $X_{\rm H} = 20$~per~cent is plotted with red dotted line, intermediate-hydrogen model with $X_{\rm H} = 30$~per~cent is demonstrated by black solid line and blue dashed line represents high-hydrogen model with $X_{\rm H} = 40$~per~cent. (a) \ion{He}{i}\,\,$\lambda6678$, (b) H\,$\alpha$ (black dashed line marks the total profile of the $X_{\rm H} = 30$~per~cent model with nebular contamination as described above), (c) \ion{Fe}{iii}\,\,$\lambda\lambda5988-6058$, (d) \ion{He}{i}\,\,$\lambda5876$, (e) \ion{N}{ii}\,\,$\lambda\lambda5666-5710$, \ion{Fe}{iii}\,\,$\lambda\lambda5712-5744$, \ion{Si}{iii}\,$\lambda5739$, \ion{Al}{iii}\,$\lambda\lambda5697,5723$, (f) \ion{Fe}{iii}\,$\lambda\lambda5073,5086,5127,5156$, (g) \ion{N}{ii}\,$\lambda\lambda4601-4643$, (h) \ion{Si}{iii}\,$\lambda\lambda4552,4567,4574$.}
    \label{fig:hydrogen}
\end{figure*}

A fundamental difficulty in determining the hydrogen abundance is created by the nebula projected onto the object. It makes a significant contribution to the observed \ion{H}{i} lines intensities. To estimate this contribution, we performed additional observations of the object (see section \ref{sect:obs}). The obtained flux-calibrated spectrum of the nebula was added to the model spectrum in absolute units with distance and interstellar extinction taken into account. We demonstrate an example of a normalized spectrum of a model with an addition of nebular lines on the panel (b) of the figure \ref{fig:hydrogen}. Nebular fluxes were normalized in order to describe the observed nebular [\ion{N}{ii}] lines.

We had created three separated series of models with different mass fractions of hydrogen: 20 (low-hydrogen), 30 (intermediate-hydrogen) and 40~per~cent (high-hydrogen). These models were created with fixed parameters of velocity distribution and clumping law to reduce the number of free parameters. We adopted the velocity law parameters as follows $\beta = 2,~v_\infty = 250~\text{km\,s}^{-1}$ and clumping law $f_1 = 0.25, f_2 = 10~\text{km\,s}^{-1}, f_3 = 40~\text{km\,s}^{-1}$. The comparison of synthetic spectra of named models is provided on the figure \ref{fig:hydrogen}.

In models with the highest hydrogen abundance, \ion{H}{i} line intensities were described at $\Dot{M} = 1.9\times10^{-5}~\text{M}_\odot\,\text{yr}^{-1}$ and $T_{\rm eff} = 19~\text{kK}$. This combination of parameters allows to reproduce the intensities of \ion{H}{i} lines but the \ion{He}{i} line intensities are underestimated. \ion{He}{i} in principle may be restored with higher $T_{\rm eff}$ but temperature-sensitive \ion{Si}{ii}\,$\lambda\lambda6347,6371$ and \ion{Fe}{iii}\,$\lambda\lambda5127,5156$ are severely underestimated with $T_{\rm eff} = 19~\text{kK}$ and would diminish further with the increase of $T_{\rm eff}$. Regarding \ion{N}{ii} lines, this model shows good agreement with \ion{N}{ii}\,$\lambda\lambda6610,5940,4895$ and \ion{N}{ii}\,$\lambda\lambda4994-5007$ lines. The intensities of \ion{N}{ii}\,$\lambda\lambda4601-4643$ emission lines, on the other hand, are underestimated in high-hydrogen model. 

Both intermediate-hydrogen model with $X_{\rm H} = 30$~per~cent and a low-hydrogen model $X_{\rm H} = 20$~per~cent describe the observed spectrum with acceptable quality. Model spectra reproduce the observed \ion{H}{i} intensities. \ion{He}{i} lines are restored within the accuracy of $10$~per~cent. Intermediate-hydrogen models underestimate this lines intensities, low-hydrogen models, conversely, overestimate them with the exception of \ion{He}{i}\,$\lambda5015$ line which is significantly better described in low-hydrogen model. Most of the \ion{N}{ii} and \ion{Fe}{iii} lines show only insignificant difference. It is worth noting the overestimation of \ion{N}{ii}\,$\lambda\lambda6610,4895$ and \ion{N}{ii},~\ion{Fe}{iii}\,$\lambda\lambda6150-6195$ lines in low-hydrogen model. It arises due to the higher $\Dot{M}$ value in comparison with the intermediate-hydrogen model (see table \ref{tab:par}). Higher $\Dot{M}$ value was required to reproduce the observed \ion{H}{i} intensities. The model with $X_{\rm H} = 30$~per~cent shows the presence of strong \ion{Al}{iii}\,$\lambda\lambda5697,5723$ and \ion{S}{iii}\,$\lambda\lambda3844,4290$ emission lines that are absent in both the observed spectrum and synthetic spectrum of the model with $X_{\rm H} = 20$~per~cent. We link this with a higher temperature of intermediate- and hugh-hydrogen models. 

\begin{table*}
\caption {The main parameters of the extended atmosphere models of AF~And. The remaining wind characteristics of the presented model are discussed in text above. $R_{\rm ph}$ is the photospheric radius defined as the radius corresponding to $\tau_{\rm Ross} = 2/3$.}
\centering
\label{tab:par}
\medskip
\begin{tabular}{l|c|c|c}
\hline
Parameter & Low-hydrogen model & Intermediate-hydrogen model & High-hydrogen model  \\
\hline
$L$, [$\text{L}_\odot$] & $6.5\times10^{5}$ & $6.5\times10^{5}$ & $6.5\times10^{5}$ \\
$\dot{M}$, $[M_{\odot}\,\text{yr}^{-1}]$ & $3.4\times10^{-5}$ & $2.7\times10^{-5}$ & $1.9\times10^{-5}$ \\
$T_*$, [kK] & $23.4$ & $ 23.4 $ & $23.4$ \\
$T_{\rm eff}$, [kK] & $17.7$ & $18.3$ & $19.2$ \\
$R_*$, [$\text{R}_\odot$] & $50$ & $50$ & $50$ \\
$R_{\rm ph}$, [$\text{R}_\odot$] & $85$ & $80$ & $73$ \\
$\beta$ & $2.0$ & $2.0$ & $2.0$ \\
$v_{\infty}$, [km\,s$^{-1}$] & 250 & 250 & 250 \\
$f_1$ & $0.25$ & $0.25$ & $0.25$ \\
$X_{\rm H}$, [\%] & 20 & 30 & 40 \\ 
\hline
\end{tabular}
\end{table*}

On the current stage of our study we cannot reliably distinguish low-hydrogen from intermediate-hydrogen model in terms of quality of agreement between the observed and model spectra.

\subsection{Dynamically-consistent wind model}
\label{sect:dynamics}
The resulting models demonstrated in section \ref{sect:hydrogen} can be used as a zeroth-order approximations for a self-consistent hydrodynamical models of the wind. The method we had used for model construction is described in details in \citet{Kostenkov2026}. Here we outline the main points of this approach.
    
The combination of the continuity equation for an isothermal spherically symmetric wind with the equation of momentum conservation for an ideal gas yields \citep{LamersCassinelli1999}
\begin{equation}\label{eom}
\left(v - \frac{a^2}{v}\right)\frac{{\rm d}v}{{\rm d}r}=\frac{2a^2}{r} - \frac{GM_*}{r^2} + g_{\rm rad}
\end{equation}

where $M_*$ is a stellar mass, $a$~--~speed of sound, $g_{\rm rad}$~--~radiative acceleration. The equation \ref{eom} has a singularity in $v(r_c) = a$. A monotonically accelerating wind is obtained under the condition $\frac{\text{d}v}{\text{d}r} > 0$ at $v(r_c) = a$ \citep{Parker1964}.

The critical point location can be found assuming the right hand side of the equation \ref{eom} is equal to zero. In our calculations we fixed the $r_c$ radius and used this equation to derive the current mass.

\begin{equation}\label{crit_point_eq}
2a^2r_{\rm c} - GM_* + g_{\rm rad}(r_{\rm c})r_{\rm c}^2=0.
\end{equation}

And with de l'H\^{o}pital rule we obtain the wind velocity derivative at the critical point \citep{LamersCassinelli1999}.

\begin{equation}\label{lhop_crit_point}
\left(\frac{{\rm d}v}{{\rm d}r}\right)_{r_{\rm c}}=\sqrt{-\frac{a^2}{r_{\rm c}^2} + \frac{GM_*}{r_{\rm c}^3} + \frac{1}{2}\left(\frac{{\rm d}g_{\rm rad}}{{\rm d}r}\right)_{r_{\rm c}}}.
\end{equation}

Finally, a complete form of a velocity distribution is found by integration of the equation \ref{eom} upwards and downwards from the critical point. Following \citet{Kostenkov2026}, we use the explicit Runge-Kutta method of 5(4) order with an adaptive step size using the Dormand-Prince formulae \citep{Dormand1980}. The velocity distribution obtained in this way does not correspond to the radiation field of the initial model. Therefore, it is necessary to conduct an iterative calculation process of several \textsc{cmfgen} models, in which the input velocity field approaches the solution of the equation of motion as it converges. The iterations are stopped when the convergence criterion is met \citep{Sundqvist2019}.
\begin{equation}\label{converge}
\max\left\vert 1  - \frac{\left(v^{\rm old} - \frac{a^2}{v^{\rm old}}\right)\frac{{\rm d}v^{\rm old}}{{\rm d}r} - \frac{2a^2}{r} + \frac{GM^{\rm old}_*}{r^2}}{g^{\rm new}_{\rm rad}} \right\vert < 1 \times 10^{-2}
\end{equation}
where the superscript `old' corresponds to the quantities of the input model and `new' stands for the updated quantity after performing the \textsc{cmfgen} model computation. 

In the previous subsection we were unable to reliably separate the $X_{\rm H} = 20$~per~cent and $X_{\rm H} = 30$~per~cent solutions. Therefore, we had constructed several self-consistent models for low and intermediate abundance of hydrogen. We were able to create self-consisted models with the main parameters ($L,~\Dot{M},~X_i$) values equal to those reported in table \ref{tab:par}. We used $r_c = 5.2 \times 10^{12}~\text{cm}$ and $r_c = 5.5 \times 10^{12}~\text{cm}$ for $X_{\rm H} = 30$~per~cent and $X_{\rm H} = 20$~per~cent models, respectively. Models with greater $r_c$ values show `colder' spectra with prominent absorption component of P~Cyg profiles of \ion{H}{i} lines and insufficient intensities of \ion{He}{i} lines. Self-consistent models with $r_c$ values lower than mentioned above are unavailable due to the arising of non-monotonic velocity law. In this case, \textsc{cmfgen} is unable to create a model.

\begin{figure}
    \centering
    \includegraphics[width=1.0\linewidth]{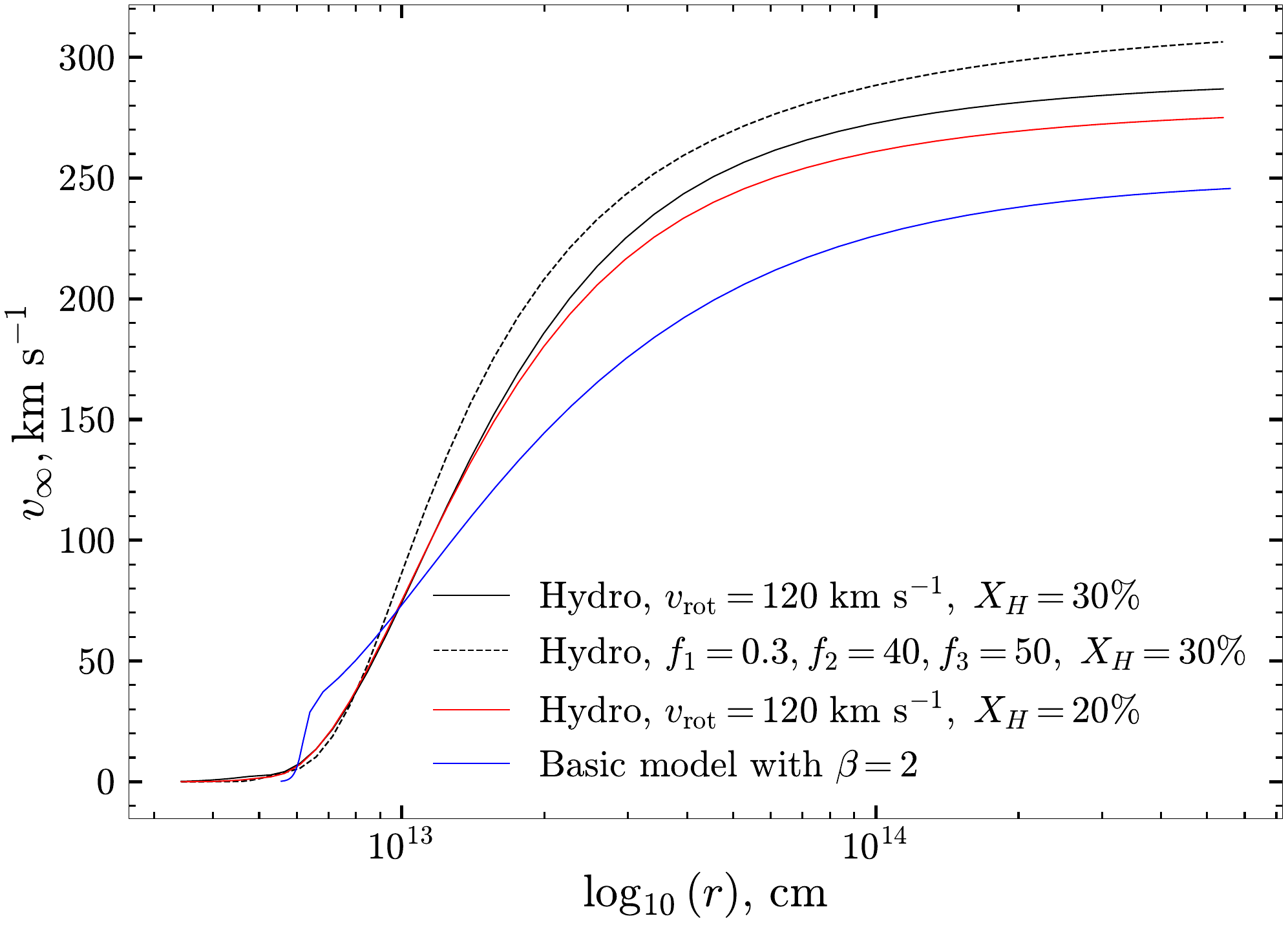}
    \caption{The forms of the wind velocity distributions for $\beta$-law based models (blue solid line) and self-consistent models discussed in the text. Black and red solid lines show the $v(r)$ for the rotating intermediate-hydrogen and low-hydrogen models, respectively. Black dashed line represents the velocity distribution of non-rotating intermediate-hydrodynamical model with $f_1 = 0.3, f_2 = 40~\text{km\,s}^{-1}~\text{and}~f_3=50~\text{km\,s}^{-1}$.}
    \label{fig:vellaw}
\end{figure}

During the construction of dynamically-consistent models we used the standard clumping law in the form of the equation \ref{clump_param} for reasons listed in the section \ref{sect:modelling}. This approach was also applied in e.g. \cite{Sander2017, Gormaz-Matamala2021}. It is important to take clumping into account when constructing dynamically-consistent models as it allows for the realistic modelling of the radiative field required for reliable radial forces calculations, and for the representation of the observed spectrum with acceptable quality. These effects are due to the enhancement of recombination rates provided by wind clumping, which increases the populations of the atomic levels responsible for line-driven wind acceleration \citep{Gormaz-Matamala2021}. Hydrodynamic models with the same $f(r)$ distribution as for $\beta$-law based models show high $v_{\infty}$ values: $v_{\infty} \approx 400~\text{km\,s}^{-1}$ for $X_{\rm H} = 30$~per~cent. This value is in strong disagreement with $v_{\infty} =250~\text{km\,s}^{-1}$ derived with our previously constructed models. The obtained stellar mass of this self-consistent model $M_* = 16.8~M_{\odot}$. We altered the adopted filling factor law to shift the minimum of $f(r)$ closer to $R_*$ to obtain lower $v_\infty$ as suggested in \citet{Kostenkov2026}. With $f_1 = 0.3, f_2 = 40~\text{km\,s}^{-1}~\text{and}~f_3=50~\text{km\,s}^{-1}$ the resulting $v_\infty = 310~\text{km\,s}^{-1}$ with close value of stellar mass. This velocity distribution is demonstrated on the figure \ref{fig:vellaw}.

The obtained model spectrum is inconsistent with the observed spectrum in terms of quality of describing \ion{N}{ii}, \ion{Si}{ii} and \ion{Fe}{iii} emission lines. \ion{H}{i} and \ion{He}{i} lines are also significantly underestimated. A model series with $X_{\rm H} = 20$~per~cent converged at $M_* = 13.5~\text{M}_\odot$ and $v_\infty = 290~\text{km\,s}^{-1}$, and overall shares described above issues with intermediate-hydrogen hydrodynamical model.

\subsection{Wind dynamics of rotating model}
\label{sect:rotation}

The insufficient quality of \ion{N}{ii} and \ion{Si}{ii-iv} lines reconstruction may be caused by e.g. temperature dependence on the polar angle in the wind. Thus, the \ion{Si}{ii} lines are reproduced better at lower temperatures ($T_{\rm eff} \lesssim 15~\text{kK}$), while the absorptions in the \ion{Si}{iii} lines are underestimated at such temperatures. Alternatively, low intensities of \ion{Si}{ii} lines may arise from a higher temperature of the outer regions of the wind. We checked this hypothesis by cooling the corresponding region with the increase of $X_N$ since nitrogen is an effective cooling agent. The observed \ion{Si}{ii} lines still show greater intensities than predicted by nitrogen-enriched ($X_N = 10~X_{N,~\odot}$) models. Returning to the first scenario, the presumed presence of regions with differing temperatures within the wind may be attributed to the high rotational velocity of the star, which constitutes a significant fraction of the critical velocity.
It is evident that the equatorial wind of a rotating star will be slower and cooler in comparison with the polar wind \citep{Maeder2003}. \citet{Levesque2025} consider the idea that the S~Dor variability observed in LBVs is triggered by rotation, which indirectly correlates with the slow rotation velocity of the `dormant' LBV P~Cygni \citep{Israelian1995} and the observed high rotation velocity of HR~Car and AG~Car (see discussion in \citealt{Groh2009HRCar}). On the other hand, the observed phenomenon of S~Dor variability may be explained by the wind-envelope interactions as suggested in \cite{Grassitelli2021} but the discussion on the physical mechanisms of LBVs variability is beyond the scope of this paper. The observed spectrum also contains a cautious indication of the possible presence of rotation in the form of the \ion{Si}{iv}\,$\lambda\lambda4088,4116$ lines, although, the latter line is contaminated by \ion{He}{i}\,$\lambda4143$. The observed width of \ion{Si}{iv}\,$\lambda4088$ exceeds the model value, which may be caused by the fact that it forms deep enough in the stellar atmosphere to be subject to the rotational broadening. Analogous effect was reported for HR~Car in \citet{Groh2009HRCar} and for AG~Car in \citet{Groh2011}.

These considerations prompted us to create a self-consistent model that includes rotation as a term in the equation of motion. Unfortunately, within the 1D radiative transfer code, we cannot accurately describe the effect of rotation on the observed line profiles, but we can obtain a realistic velocity law in this case. According to \citet{LamersCassinelli1999}, for the rotating star the momentum equation \ref{eom} has an additional term due to centrifugal force:
\begin{equation}\label{eom_rot}
\left(v - \frac{a^2}{v}\right)\frac{{\rm d}v}{{\rm d}r}=\frac{2a^2}{r} - \frac{GM_*}{r^2} + g_{\rm rad} + \frac{v_{\rm rot}^2 R^2_*}{r^3}
\end{equation}
where $v_{\rm rot}$ is the rotation velocity in the equatorial plane at $R_*$.
Next, equations \ref{crit_point_eq}-\ref{converge} must be rewritten to account for the new term. The critical point equation becomes
\begin{equation}\label{crit_point_eq_rot}
2a^2r_{\rm c} - GM_*  + g_{\rm rad}(r_{\rm c})r_{\rm c}^2 + \frac{v_{\rm rot}^2 R^2_*}{r_{\rm c}} = 0
\end{equation}

Next, the velocity derivative transforms into
\begin{equation}\label{lhop_crit_point_rot}
\left(\frac{{\rm d}v}{{\rm d}r}\right)_{r_{\rm c}}=\sqrt{-\frac{a^2}{r_{\rm c}^2} + \frac{GM_*}{r_{\rm c}^3} + \frac{1}{2}\left(\frac{{\rm d}g_{\rm rad}}{{\rm d}r}\right)_{r_{\rm c}} - \frac{3}{2}\frac{v_{\rm rot}^2R_*^2}{r_{\rm c}^4}}
\end{equation}

And, finally, the convergence criterion
\begin{equation}\label{converge_rot}
\max\left\vert 1  - \frac{\left(v^{\rm old} - \frac{a^2}{v^{\rm old}}\right)\frac{{\rm d}v^{\rm old}}{{\rm d}r} - \frac{2a^2}{r} + \frac{GM^{\rm old}_*}{r^2} - \frac{v_{\rm rot}^2 R_*^2}{r^3}}{g^{\rm new}_{\rm rad}} \right\vert < 1 \times 10^{-2}
\end{equation}

The solution to the equation of motion \ref{eom_rot} ($r_c = 6.9 \times 10^{12}~\text{cm}$ for $X_{\rm H} = 30$~per~cent) yielded a velocity distribution with $v_\infty \approx 290~\text{km\,s}^{-1}$. The velocity profile of this self-consistent model is shown on the figure \ref{fig:vellaw}. The current mass of the star $M_* = 17.7~\text{M}_\odot$. We tested several $v_{\rm rot}$ values and adopted $v_{\rm rot} = 120~\text{km\,s}^{-1}$. With this approach we have succeeded in the creation of a model with a slightly higher $\Dot{M} = 3\times10^{-5}~\text{M}_\odot\,\text{yr}^{-1}$. This value allows for a better description of the observed intensities of \ion{H}{i} and \ion{He}{i} lines. Further increase of $\Dot{M}$ in hydrodynamical models leads to a decrease of the wind velocity and effective temperature.

\begin{figure*}
    \centering
    \includegraphics[width=0.9\linewidth]{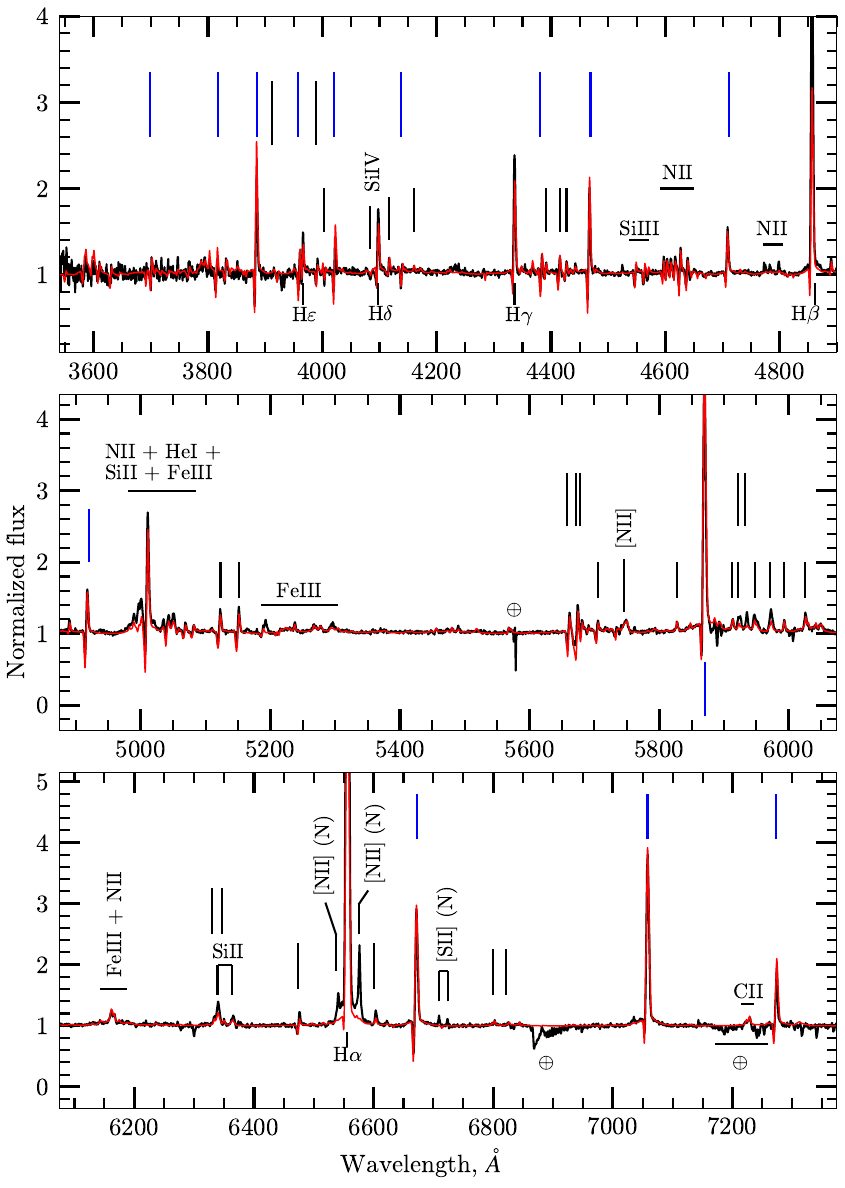}
    \caption{Comparison between the observed spectrum of AF~And (black solid line) and the low-hydrogen rotating hydrodynamical model (red solid line). The model is shifted to the observed wavelengths and convolved with instrumental profile for clarity. \ion{He}{i} lines are indicated with blue vertical lines, \ion{N}{ii} lines are shown with long black vertical lines, short black vertical lines denote \ion{Fe}{iii} lines, $\oplus$ sign marks the telluric features. Nebular lines are labelled with `(N)' sign.}
    \label{fig:fullspec}
\end{figure*}

The low-hydrogen model series shows similar behaviour with almost the same location of critical point as for models with $X_{\rm H} = 30$~per~cent ($r_c = 6.7 \times 10^{12}~\text{cm}$) and the same $v_{\rm rot} = 120~\text{km\,s}^{-1}$. This model series converged at $M_* = 15.6~\text{M}_\odot$ and $v_\infty = 275~\text{km\,s}^{-1}$. Here we do not increase $\Dot{M}$ value any further since the model would have lower $v_\infty$ value. Within the form of the self-consistent velocity distribution this would lead to an underestimation of blue-shifted absorption components of P~Cyg profiles of \ion{He}{i} lines. We provide the comparison between the synthetic and the observed spectra on the figure \ref{fig:fullspec}. The resulting model parameters are listed in table \ref{tab:par_rot}.

\begin{table}
\caption {The main parameters of the hydrodynamical models of AF~And with rotation taken into account. $R_{\rm ph}$ is the photospheric radius defined as the radius corresponding to $\tau_{\rm Ross} = 2/3$.}
\centering
\label{tab:par_rot}
\medskip
\begin{tabular}{l|c|c}
\hline
Parameter & Low-hydrogen model & Intermediate-hydrogen model \\
\hline
$L$, [$\text{L}_\odot$] & $6.5\times10^{5}$ & $6.5\times10^{5}$ \\
$\dot{M}$, $[M_{\odot}\,\text{yr}^{-1}]$ & $3.4\times10^{-5}$ & $3.0\times10^{-5}$ \\
$M_*$, [$\text{M}_\odot$] & $15.6$ & $ 17.7 $ \\
$T_*$, [kK] & $20.4$ & $ 21.4 $ \\
$T_{\rm eff}$, [kK] & $16.1$ & $16.1$ \\
$R_*$, [$\text{R}_\odot$] & $64$ & $60$ \\
$R_{\rm ph}$, [$\text{R}_\odot$] & $104$ & $104$ \\
$v_{\infty}$, [km\,s$^{-1}$] & 275 & 290 \\
$f_1$ & $0.1$ & $0.1$ \\
$X_{\rm H}$, [\%] & 20 & 30 \\ 
\hline
\end{tabular}
\end{table}

Model spectra of both low- and intermediate-hydrogen models show a relatively accurate profiles of the most prominent spectral features of the observed spectrum of AF~And such as \ion{H}{i}, \ion{He}{i}, \ion{N}{ii} and \ion{Fe}{iii} intensities. Rotating hydrodynamical models describe the \ion{Si}{ii-iii} lines profiles more precisely than the regular $\beta$-law based models. However, it is worth noting that \ion{Si}{iv}, \ion{N}{ii}\,$\lambda\lambda4987-5010$ and \ion{Fe}{iii}, \ion{N}{ii}\,$\lambda\lambda5918-5956$ lines are underestimated in self-consistent models. We link this effect with a lower temperatures of these models but simultaneously remark that the temperature-inhomogeneous wind of the rotating star could possibly describe the observed spectrum. 

\section{Discussion}
\label{sect:discussion}

\subsection{Wind properties}

The form of the velocity law of AF~And in the observed state was mainly controlled by \ion{Fe}{iii} transitions (up to 56~per~cent in outer regions above critical point). Other ions of the iron group elements also provide a significant contribution in total wind acceleration. In the inner part of the wind, however, the most effective accelerating ion is \ion{Fe}{iv}. In the case of AF~And, this behaviour of iron ionization structure obtained during modelling does not lead to the two-component form of the velocity law, which is observed in the hotter ($T_{\rm eff} \approx 24~\text{kK}$) LBV Var\,2 \citep{Kostenkov2026}. In the presented models, the recombination of \ion{Fe}{iv} to \ion{Fe}{iii} occurs in the vicinity of the critical point, which, according to \citet{Vink1999}, corresponds to AF~And being located near the bi-stable limit. Although, crossing the bi-stable limit is not physically related to the transition towards WR stars \citep{Meynet2005}, it is still interesting to compare the properties of the wind of AF~And with those of O- and WR stars. \citet{Vink2012} proposed a relationship between the wind efficiency parameter $\eta$ and its optical depth $\tau_w$, $\eta = \frac{\Dot{M} v_\infty}{L/c} = f_{\rm corr} \tau_w$. For AF~And, $\eta \approx 0.7$. It is expected that for O stars $\eta < 1$ \citep{Repolust2004}, and for WR stars, conversely, $\eta > 1$ \citep{Nugis2000}. Both rotating models give the optical depth of the AF~And wind $\tau_w = \int^{\infty}_{r_c} \kappa \rho \text{d}r \approx 2$ which leads to $f_{\rm corr} \approx 0.3$ in the case of AF~And. This value is lower than the typical $f_{\rm corr} \approx 0.6$ reported in \citet{Vink2012} for the most massive stars of the Arches cluster. This paper also defines the transitional mass-loss rate for the transition state between O- and WR-like winds that occurs at $\tau_w = 1$. Adopting optical depth, obtained for AF~And, the transitional $\Dot{M} = 1.4\times10^{-5}~\text{M}_\odot\,\text{yr}^{-1}$ which is considerably lower than our estimate obtained by \textsc{cmfgen} modelling. This consideration places the wind of AF~And closer to WR-like state, however, the typical temperature of the WNh stars (see table 2 in \citealt{Hamann2006}) is much higher than derived for AF~And. 

\subsection{Wind inhomogeneities}
\label{sect:clumping}

Both self-consistent and regular models we demonstrate in this paper are in good agreement with observed spectrum of AF~And. However, there are some discrepancies between them in terms of intensities of some emission lines. \ion{Si}{iv}\,$\lambda4088,4116$, \ion{N}{ii}\,$\lambda\lambda4987-5010$ and \ion{Fe}{iii}, \ion{N}{ii}\,$\lambda\lambda5918-5956$ lines are better represented in the basic model with $\beta = 2$. This effect may arise due to underestimation of radiative acceleration of the wind in hydrodynamical model which leads to a colder atmospheres of these models. This consideration encouraged us to study the influence of the wind inhomogeneities distribution law since it affect resulting velocity profiles \citep{Bernini-Peron2026, Kostenkov2026}.

\begin{figure*}
    \centering
    \includegraphics[width=1.0\linewidth]{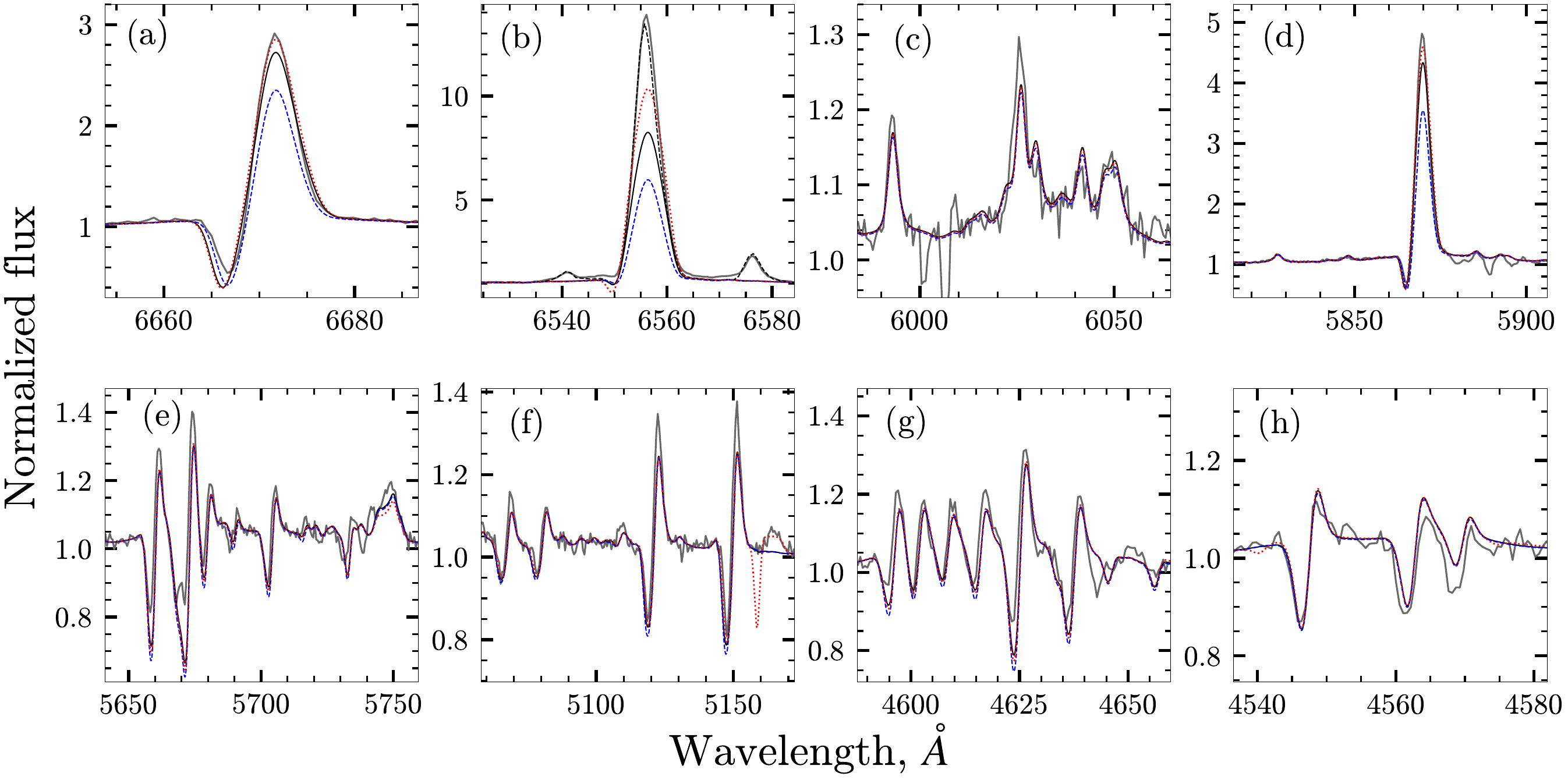}
    \caption{Comparison of the lines profiles for several chosen lines in dynamically-consistent models with different $f(r)$ laws. The normalized spectrum of AF~And is shown by grey solid line, the exponential law with $f_1 = 0.1, f_2 = 70~\text{km\,s}^{-1}$ is plotted with red dotted line, the model with $f(r)$ minimum at $10~R_*$ is demonstrated by black solid line and blue dashed line represents the spectrum of the model with $f(r)$ minimum at $2~R_*$. The demonstrated line set is the same as in the figure \ref{fig:hydrogen}. The black dashed line on the panel b) shows the combined model and nebular lines profiles as on the figure \ref{fig:hydrogen}.}
    \label{fig:clumping}
\end{figure*}

Observations and modelling of inhomogeneities in the wind indicate a possible minimum of the filling factor at a radius approximately equal to $2~R_*$ \citep{Puls2006, Najarro2011, Sundqvist2013}. On the other hand, some studies devoted to radiation-hydrodynamic calculations of the wind structure show that $f(r)$ reaches its minimum significantly further away, at distances of $10-20~R_*$ \citep{Runacres2007, RubioDiez2022}.
We tested the impact of the choice of $f(r)$ distribution parameters on the synthetic spectrum and the shape of the velocity law. Several rotating models with $X_{\rm H} = 30$~per~cent and different $f(r)$ distributions had been created since model spectra of low- and intermediate-hydrogen models are practically indistinguishable. We also compared the resulting models with a model that uses a simpler exponential distribution of the filling factor ($f_3 = 0$). When constructing the models, our aim was to track the behaviour of the $f(r)$ distribution shape. To this end, the $f_1,~f_2,~f_3$ parameters were selected to ensure the same $f(r_c)$ value in each model, since the balance of radial forces at the critical point depends on the wind density and determines the shape of the velocity law. The other parameters values were equal to those presented in the table \ref{tab:par_rot}.

The results are presented in figure \ref{fig:clumping}. We show that our models with $f(r)$ minimum at $2~R_*$ are less suitably describe the observed spectrum. A suitable agreement between the observed spectrum and the model was achieved for both models with a $f(r)$ minimum near $10~R_*$ and with exponential clumping law. The latter model shows higher intensities in \ion{H}{i}, \ion{He}{i} and \ion{Si}{ii} lines compared to the former, but shows the presence of \ion{Fe}{ii}\,$\lambda5169$ line with a prominent absorption component that is absent in the observed spectrum. Due to the absence of IR data we are unable to reliably establish the exact form of $f(r)$ distribution, therefore we conclude that the minimum of inhomogeneities distribution is unlikely to be located in the close vicinity of the hydrostatic radius $R_*$. 

We also investigated the influence of optically thick clumps on the wind dynamics, using so-called `macroclumping' approximation. According to \citet{Hamann2008} this approach assumes that the clumps have a uniform size and are statistically distributed with the separation parameter $L$. The presence of macroclumps changes the wind opacity with respect to the relations presented in \citet{Hamann2008}. We used the updated effective opacity to modify $g_{\rm rad}$ obtained from the \textsc{cmfgen} model for further use in the equation of motion \ref{eom}. The clump separation law can be parametrized as

\begin{equation}
\label{mc_param}
    L(r) = L_0 \left(r^2 \frac{v(r)}{v_\infty} \right)^{1/3}
\end{equation}
with $L_0$ is the typical clump separation in the units of stellar radius.

We had calculated several intermediate-hydrogen non-rotating models (parameters are listed in the table \ref{tab:par}) with different values of $L_0$ and found out that models with $L_0 = 0.05$ give the velocity distribution close to the one obtained for the rotating model. This model series has $v_\infty \approx 320~\text{km\,s}^{-1}$ and a stellar mass $M_* \approx 17.7~\text{M}_\odot$. This result shows that macroclumping can significantly change the shape of the wind velocity law, in particular, we were able to create a model with a velocity distribution that replicates that of the rotating model with an accuracy of about 10~per~cent. Macroclumping could also significantly change the line profiles. \citet{Petrov2014} report on the high sensitivity of H\,$\alpha$ line profile to the presence of the optically thick clumps. However, we abstain from any further analysis of macroclumping since we can only investigate the behaviour of the velocity distribution as \textsc{cmfgen} can not handle the presence of optically thick inhomogeneities during radiative transfer calculations.

\subsection{Evolutionary status}
\label{sect:evol}

In order to determine the evolutionary status of AF~And we had utilized the grids of the \textsc{GENEC} evolutionary models of Geneva group \citep{Ekstrom2012} for solar metallicity. Evolutionary models show that the rotation velocity of the star significantly decreases during its lifetime. For example, the model with $M_{\rm init} = 40~\text{M}_\odot$ and $v_{\rm rot,~init} = 0.4~v_{\rm crit}$ at stage with surface hydrogen abundance $X_{\rm H,~surf} = 30$~per~cent has $v_{\rm rot} = 0.05~\text{km\,s}^{-1}$. Since the rotating evolutionary tracks predict  negligible rotation velocity at the advanced evolutionary stages, we used both rotating and non-rotating tracks for this study.

\begin{figure}
    \centering
    \includegraphics[width=0.9\linewidth]{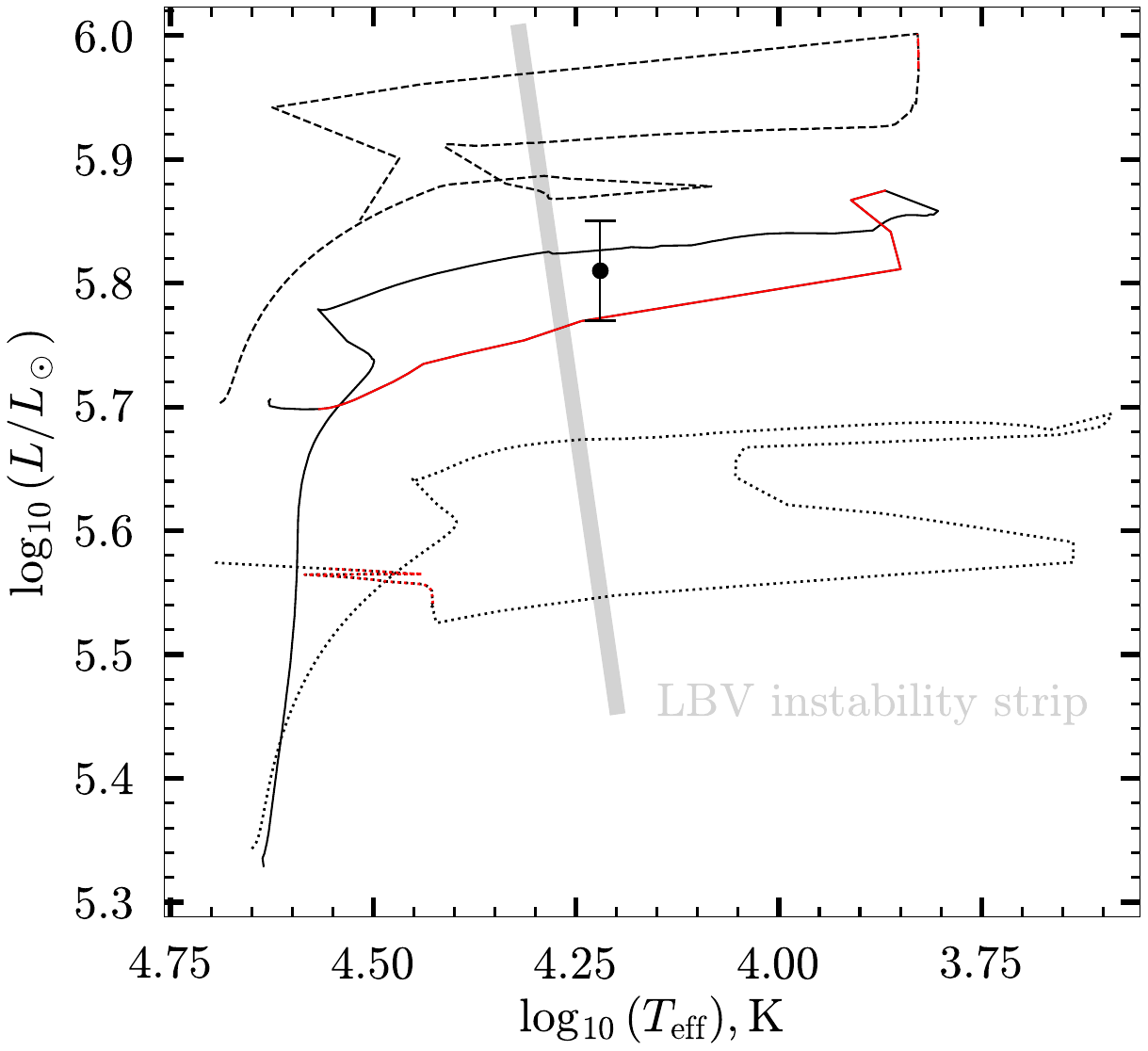}
    \caption{The location of AF~And (marked with black circle) in the hot state on Hertzsprung-Russel diagram in comparison with the Geneva group evolutionary tracks for solar metallicity. Dotted line represents the evolutionary track for non-rotating star with initial mass of $40~\text{M}_\odot$, solid line demonstrates the evolutionary track for rotating ($v_{\rm rot} = 0.4~v_{\rm crit}$) star with initial mass of $40~\text{M}_\odot$ and dashed line shows the the evolutionary track for non-rotating star with initial mass of $60~\text{M}_\odot$. The luminosity error bar accounts for statistical uncertainties of the measurements of magnitudes, interstellar extinction (see section \ref{sect:obs}) and distance values from \citet{McConnachie2005}. Red colour indicates the $20 <X_{\rm H,~surf} < 40$~per~cent stage. The instability strip is adapted from \citet{Groh2009AGI}.}
    \label{fig:tracks}
\end{figure}

The figure \ref{fig:tracks} demonstrates the location of AF~And on the Hertzsprung-Russell diagram for the luminosity and effective temperature values of $L = 6.5\times10^5~\text{L}_\odot$ and $T_{\rm eff} = 16~\text{kK}$, respectively. Present model grid does not contain any models that exactly match the observed parameters derived with our modelling. The evolutionary track for rotating star with $M_{\rm init} = 40~\text{M}_\odot$ lies close enough in terms of $L$, but current mass of the star $M_* = 32~\text{M}_\odot$ significantly higher than estimate we derived from hydrodynamic modelling. $X_{\rm H,~surf}$ is also moderately higher, 40 versus 30~per~cent from \textsc{cmfgen} modelling. It is worth pointing out that for $M_* = 32~\text{M}_\odot$ evolutionary model gives $\Dot{M} = 2.4\times10^{-5}~\text{M}_\odot\,\text{yr}^{-1}$ which is comparable with our results. Non-rotating track for the same initial mass shows significantly lower $L$ values. On the other hand, it shows that surface hydrogen depletion occurs on lower stellar mass values (e.g. $X_{\rm H,~surf} = 30$~per~cent for $M_* = 15.6~\text{M}_\odot$). If we consider the evolutionary track of a non-rotating, more massive star with $M_{\rm init} = 60~\text{M}_\odot$, we find that it will have exhausted almost all the hydrogen on its surface ($X_{\rm H,~surf} = 13$~per~cent) by the time it reaches the age of $t = 3.6$~Myr. At that point, it will have a mass of $M_* \approx 26~\text{M}_\odot$, a high luminosity $L \approx 9\times 10^5~\text{L}_\odot$ and a temperature of about $15~\text{kK}$. Upon reaching $17.7~\text{M}_\odot$, there will be no hydrogen left on the surface and the star will have a luminosity of $L \approx 4.4 \times10^5~\text{L}_\odot$ and $T_{\rm eff} \approx42~\text{kK}$. We note than non-rotating evolutionary tracks predict a much shorter phase with $X_{\rm H,~surf} = 20 - 40$~per~cent typical for LBVs.

Taking these considerations into account, we estimate the initial mass of AF~And to be $M_{\rm init} = 43-53~\text{M}_\odot$, depending on its initial rotation velocity, and the current age of the star to be $t \approx 5~\text{Myr}$. A higher initial mass would have allowed for higher luminosity, while surface hydrogen exhaustion would occur at lower masses. The high current rotation velocity, however, is not supported by evolutionary models and may indicate that AF~And have spent part of its evolutionary history in a binary system.

\section{Conclusion}
\label{sect:conclusion}

In addition to specific applications of our results for further study of AF~And, the properties of this object and the methods employed may be useful for obtaining information about the evolutionary status of stars of this class, thanks to the possibility of mass estimation from dynamically consistent modelling. The hydrodynamic calculations performed allow us to depart from the widely used $\beta$-law, which will reduce the number of free parameters in the modelling and simultaneously allows to obtain more realistic velocity laws that take into account the balance of radial forces in the winds.

The direct solution of the equation of motion for stellar wind yields the non-parametric velocity distribution which is influenced by clumping law, among the other input parameters of the model. The importance of clumping parameters choice in context of its influence on the velocity field is demonstrated in the section \ref{sect:dynamics}. However, the inhomogeneities distribution in the winds of LBVs remains unclear. Therefore, we are compelled to apply the standard approach based on the studies of other classes of massive stars. Thus, the relatively high number of free parameters in the presented models of AF~And inherently limits the accuracy of the fundamental parameter estimation. In this regard, the question of a possible degeneracy between the clumping distribution and other fundamental parameters arises. In the case of AF~And, a significant impact on parameter estimation is not expected, considering the results of the section \ref{sect:clumping}. To confirm or refute the existence of this issue in general, large grids of dynamically-consistent models in a multidimensional parameter space (preferably all fundamental parameters of the models) must be created. Such vital research is, however, far beyond the scope of this work.

With our modelling we showed that the hydrogen abundance of AF~And is more likely to be in the range of $20 - 30$~per~cent which may indicate a more advanced evolutionary stage of AF~And compared to other LBVs. It was also found that another stellar and wind parameters are almost independent of the hydrogen abundance choice with the exception of the stellar mass. We estimate it to be $M_* = 15.6~\text{M}_\odot$ and $M_* = 17.7~\text{M}_\odot$ for $X_{\rm H} = 20$~per~cent and $X_{\rm H} = 30$~per~cent, respectively.

\section*{Acknowledgements}

We thank D.J.~Hillier for making \textsc{cmfgen} code publicly available. We are grateful to A.V.~Moiseev for the opportunity to use SCORPIO-2 focal reducer in our study. The authors thank an anonymous referee for valuable comments that improved this manuscript.
Observations with the SAO RAS telescopes are supported by the Ministry of Science and Higher Education of the Russian Federation. The renovation of telescope equipment is currently provided within the national project `Science and universities'.
This work made use of Astropy:\footnote{http://www.astropy.org} a community-developed core Python package and an ecosystem of tools and resources for astronomy.
This research used the facilities of the Canadian Astronomy Data Centre operated by the National Research Council of Canada with the support of the Canadian Space Agency (\url{https://www.cfht.hawaii.edu/en/science/Publications/index.php#ACK}).
This study was supported by the Russian Science Foundation, project \textnumero~26-12-00263 `The nature of extremely luminous stars: bridging observational statistics and physical models',  https://rscf.ru/project/26-12-00263/.

%%%%%%%%%%%%%%%%%%%%%%%%%%%%%%%%%%%%%%%%%%%%%%%%%%
\section*{Data Availability}

The data underlying this article will be shared on reasonable request to the corresponding author.

%%%%%%%%%%%%%%%%%%%% REFERENCES %%%%%%%%%%%%%%%%%%

% The best way to enter references is to use BibTeX:

\bibliographystyle{mnras}
\bibliography{bibliography} % if your bibtex file is called example.bib

% Alternatively you could enter them by hand, like this:
% This method is tedious and prone to error if you have lots of references
%\begin{thebibliography}{99}
%\bibitem[\protect\citeauthoryear{Author}{2012}]{Author2012}
%Author A.~N., 2013, Journal of Improbable Astronomy, 1, 1
%\bibitem[\protect\citeauthoryear{Others}{2013}]{Others2013}
%Others S., 2012, Journal of Interesting Stuff, 17, 198
%\end{thebibliography}

%%%%%%%%%%%%%%%%%%%%%%%%%%%%%%%%%%%%%%%%%%%%%%%%%%

%%%%%%%%%%%%%%%%% APPENDICES %%%%%%%%%%%%%%%%%%%%%

%\appendix

%\section{Some extra material}

%If you want to present additional material which would interrupt the flow of the main paper,
%it can be placed in an Appendix which appears after the list of references.

%%%%%%%%%%%%%%%%%%%%%%%%%%%%%%%%%%%%%%%%%%%%%%%%%%

% Don't change these lines
\bsp	% typesetting comment
\label{lastpage}
\end{document}